\documentclass[12pt,article]{revtex4-1}
\usepackage{amsmath}

\usepackage{graphicx}
\usepackage{dcolumn}
\usepackage{bm}
\usepackage{amssymb}   % for math
\usepackage{amsmath}
\usepackage{epsfig}
\usepackage[T1]{fontenc}
\usepackage{appendix}
\usepackage{color}

\begin{document}

\title{Reversibility of laser filamentation}

\author{N. Berti,$^1$ W. Ettoumi,$^1$ J. Kasparian,$^{2,\ast}$ and J.-P. Wolf$^1$ }
\address{1. Universit\'e de Gen\`eve, GAP-Biophotonics, Chemin de Pinchat 22, 1211 Geneva 4, Switzerland\\
2. Universit\'e de Gen\`eve, GAP-Nonlinear, Chemin de Pinchat 22, 1211 Geneva 4, Switzerland}

\email{jerome.kasparian@unige.ch}

\begin{abstract}
We investigate the reversibility of laser filamentation, a self-sustained, non-linear propagation regime including dissipation and time-retarded effects. We show that even losses related to ionization marginally affect the possibility of reverse propagating ultrashort pulses back to the initial conditions, although they make it prone to finite-distance blow-up susceptible to prevent backward propagation.
\vspace*{4pt}
\end{abstract}

\date{\today}

\maketitle

%\ocis{190.0190 Nonlinear optics; 190.3270 Kerr effect; 190.7110 Ultrafast nonlinear optics; 190.5940 Self-action effects; 190.3100 Instabilities and chaos; 190.5530 Pulse propagation and solitons.}

%\pacs{42.65.Jx Beam trapping, self focusing and defocusing,self-phase modulation; 42.65.Tg Optical solitons; 02.30.Zz Inverse problems; 11.30.Er Charge conjugation, parity, time reversal, and other discrete symmetries.}

%\vspace*{4pt}

\section{Introduction}

%Deterministic chaos clearly highlights that determinism does not ensure the time invertibility of physical systems~\cite{Lorenz}. 
Non-linear systems give rise to attractors, relaxation, dissipation, as well as couplings, both external  and internal, that can induce a loss of memory of the initial conditions within a finite time, preventing from recovering the initial conditions by reversing time in the dynamics. While the invertibility of linear differential systems is well understood~\cite{BenArtzi,Chernyshov}, its non-linear counterpart remains an open question in many areas of non-linear physics  and mathematics~\cite{Fliess}.

Optical systems provide a well-suited framework to study invertibility because of the availability of efficient numerical models, as well as the possibility to validate them with tabletop experiments in a wide variety of experimental parameters. 
Time invertibility has been demonstrated in several non-linear optical systems (optical fibers~\cite{Tsang2003}, solid focusing~\cite{Goy} and defocusing~\cite{BarsiWF2009} Kerr media, coupled-resonator optical waveguides~\cite{Wang2012}), where it is generally termed as time-reversibility. In this work, we shall stick to this common terminology, although strictly speaking reversibility is a stronger condition as it depicts systems with full inversion symmetry with regard to time, like, e.g. a pendulum.

Here, by numerically investigating laser filamentation and the associated ionization as well as molecular alignment, we show that time reversibility can be maintained even in this three-dimensional, lossy non-linear system. Filamentation~\cite{BraunKLDSM1995,CouaironM07,Chin,Kasparian,Gavish} is a propagation regime typical of high-power, ultrashort-pulse lasers. It stems from a dynamic balance between Kerr self-focusing and self-defocusing non-linearities of higher orders, including  the laser-generated plasma and the saturation of the medium polarisability under strong-field illumination~\cite{PopovJETP,IvanovPNAS,IvanovNJP,BejCHLKWF2013}.

Due to this dynamic balance, the intensity tends to relax to a so-called clamping intensity ($\approx$ 60 TW/cm$^2$ in air at 800 nm) \cite{KaspaSC2000}. As this clamping intensity was shown to be pretty independent from the initial conditions \cite{DaiglJHWKRBC2010}, it could be expected to act as an attractor and induce a loss of memory of the initial conditions during the propagation. Similarly, the self-healing \cite{CourvBKSMYW2003,LiuTAGBC2005} and mode-cleaning \cite{PradeFMCBEKSV2006} properties of filamentation imply the convergence of the system towards a stable mode that appears independent of the initial conditions, at least if one observes the intensity. Finally, the extremely high dephasing at the non-linear focus implies a large cumulative nonlinear phase shift that is highly sensitive to small fluctuations of the initial power, resulting in the loss of the phase information when measurements are averaged over several pulses \cite{Gaeta_phase}.

Here, we demonstrate numerically that, despite these indications of memory loss, laser filamentation is time-reversible with minimal error. We were able to recover accurately the initial conditions by back-propagating the final electric field obtained at the end of laser filament propagation, without any further additional information, in particular without the need to store the free electron concentration along the propagation pathway. 

\section{Reversibility of the cubic-quintic non-linear Schr\"odinger Equation}

If considering a stable structure in the co-propagating frame travelling at the pulse group velocity, reverse-propagating the system requires to exchange $z$ and $-z$, which in the moving frame is equivalent to inverting the global time. 
Let us apply this approach to the non-linear Schr\"odinger equation  (NLSE) in a general form for the complex field envelope of an electric field $\varepsilon(r,z)$:
\begin{equation}
\mathrm{i}\partial_z \varepsilon + \Delta_\bot \varepsilon + f(\left|\varepsilon\right|^2)\varepsilon=0,
\label{eqn:NLS_CQ}
\end{equation}
where the operator $\Delta_\bot$ is the transverse Laplacian ($r^{-1} \partial_r r \partial_r$ up to a normalization constant in a cylindrical symmetry). For example, if one sets $f(|\varepsilon|^2)=\gamma|\varepsilon|^2-\delta|\varepsilon|^4$, this equation describes the paraxial propagation of a linearly polarized laser beam in a cubic-quintic nonlinear medium \cite{BergeCQ}. As long as $f$ only depends on $|\varepsilon|^2$ (i.e., on the intensity and not directly on the complex electric field $\varepsilon$), reverse propagating the system by changing $z$ to $-z$, yields
\begin{equation}
\mathrm{i}\partial_z \varepsilon^\ast + \Delta_\bot \varepsilon^\ast + f(\left|\varepsilon\right|^2)\varepsilon^\ast=0,
\label{eqn:NLS_CQ_inverse}
\end{equation}
Changing $\varepsilon$ to its complex conjugate $\varepsilon^\ast$ yields back Equation (\ref{eqn:NLS_CQ}). The backward propagation of a cubic-quintic system is therefore equivalent to the forward propagation, since the real electrical field is proportional to $\varepsilon+\varepsilon^\ast$, so that $\varepsilon$ and $\varepsilon^\ast$ play equivalent roles. Consequently, one can recover \emph{any} initial observable starting from any propagated solution by simply integrating Eq.~(\ref{eqn:NLS_CQ}). 

However, the same reasoning cannot be applied when energy dissipation is taken into account. If we  introduce a linear loss term in the original NLSE:
\begin{equation}
\mathrm{i}\partial_z \varepsilon + \Delta_\bot \varepsilon + f(\left|\varepsilon\right|^2)\varepsilon + \mathrm{i}\delta \varepsilon=0.
\label{eqn:NLS_linear_dissipation}
\end{equation}
Switching $z$ to $-z$ yields the reverse propagating equation, which can be expressed in terms of the conjugate field $\varepsilon^\ast$ as
\begin{equation}
\mathrm{i}\partial_z \varepsilon^\ast + \Delta_\bot \varepsilon^\ast + f(\left|\varepsilon\right|^2)\varepsilon^\ast - \mathrm{i}\delta \varepsilon^\ast=0,
\label{eqn:NLS_linear_dissipation_backwards}
\end{equation}
where the former dissipation term now acts as a gain. A simple variance analysis~\cite{Fibich2001} shows that the backward equation is unstable. In some initial conditions on $\varepsilon^\ast$, it can blow-up at a finite distance. As a consequence, backward propagation of a generic lossy NLSE is a priori an ill-posed problem highly sensitive to the amplification of noise, whether from numerical error or from the physical system itself.

\begin{figure}[t!]
	\centerline{
		\includegraphics[width=1.0\columnwidth]{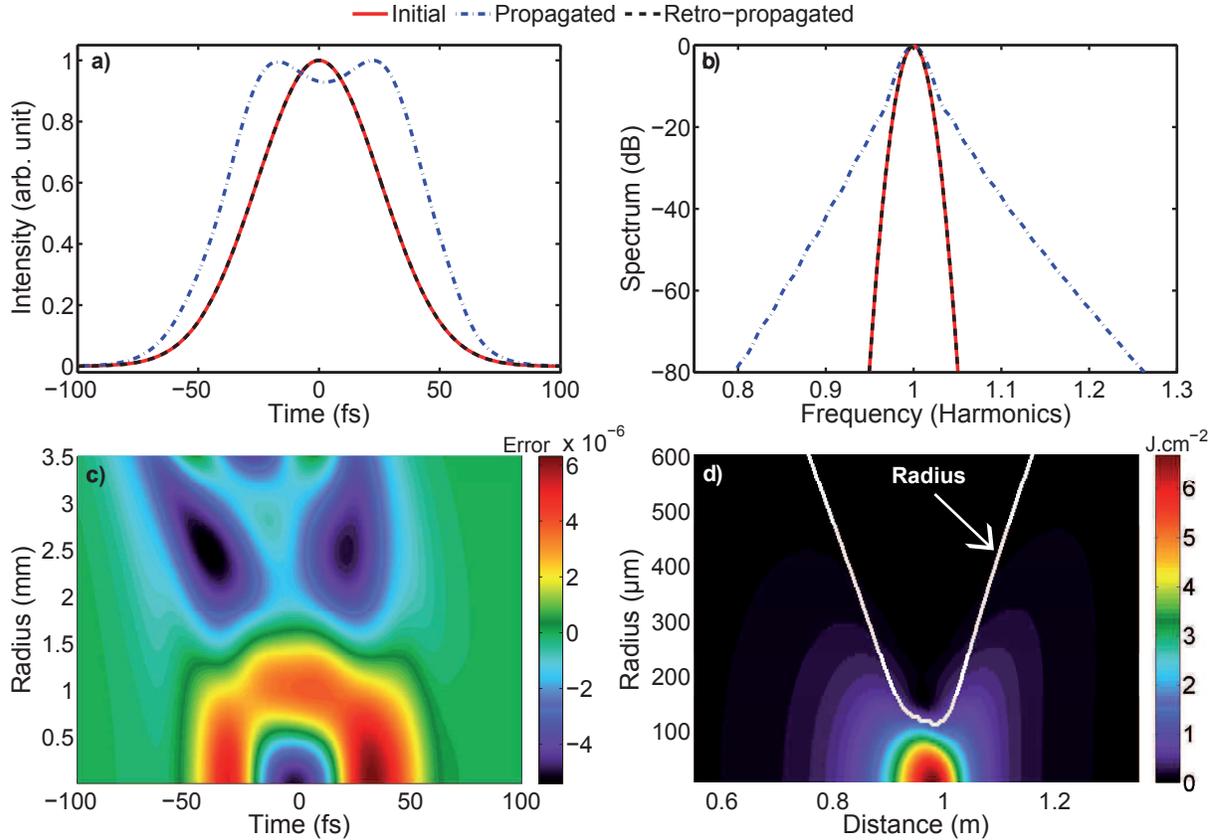}
		}
		\caption{Back-propagation of a laser filament within the frame of the cubic-quintic model. (a) On-axis intensity and (b) spectrum of the initial (red solid line), propagated (blue dash-dotted line) and back-propagated pulses (black dashed line); (c) Relative error on intensity between the initial and back-propagated pulses; (d) Fluence distribution along the propagation distance. The solid white line indicates the beam radius.}
	\label{CubQuintic_fig}
\end{figure}

To be able to assess for the impact of the non-reversible terms of the description of filamentation, we will first evaluate the magnitude of the numerical errors due to our {"forth and back"} integration scheme and implementation by using the formally reversible cubic-quintic case (Equation (\ref{eqn:NLS_CQ})) as a benchmark.

In practice, for more completeness, we numerically integrated the Unidirectional Pulse Propagation Equation (UPPE)~(\ref{eq_CubQ}) in the approximation of cylindrical symmetry~\cite{KolesMM2002,Kolesik2004}.  Its reversibility can be established exactly in the same way as for the NLSE.
Let us consider for that purpose the real electric field $\overrightarrow{E}(r,z,t)=\frac{1}{2}[\varepsilon(r,z,t)+\varepsilon^{*}(r,z,t)]\vec{u}_\perp$, linearly polarized along $\vec{u}_\perp$, which is orthogonal to the propagation direction $\vec{u}_{z}$. The propagation equation then reads:
\begin{equation}
\partial_{z}\widetilde{\varepsilon}= \textrm{i}(k_{z}-\dfrac{\omega}{v_\mathrm{g}}) \widetilde{\varepsilon} + \dfrac{\textrm{i}\omega^{2}}{c^{2}k_{z}}  \sum_{j=1}^{2}  n_{2j} \widetilde{|\varepsilon|^{2j}\varepsilon},
\label{eq_CubQ}
\end{equation}
in which the electrical field $\varepsilon(r,z,t)$ now also depends on {local} time, and the tilde denotes a Fourier-Hankel transform
\begin{equation}
\widetilde \varepsilon(k_{\bot},z,\omega)=\iint r \mathrm{J}_{0}(k_{\bot}r) \varepsilon(r,z,t)\mathrm{e}^{\mathrm{i}\omega t} \mathrm{d}t \mathrm{d}r.
\end{equation}
Here $k_{\bot}$ is the transverse wavevector, J$_0$ the zeroth-order Bessel function, $v_\mathrm{g}$ the group velocity, $\epsilon_{0}$ the vacuum permittivity, $c$ the speed of light, and $k_{z}=\sqrt{{k}^{2}(\omega)-{k}_{\bot}^{2}}$ with $k(\omega)=n(\omega)\omega/c$, $n(\omega$) being the refractive index at frequency $\omega$, $n_{2}=1.0\times10^{-23}$ m${^2}$.W$^{-1}$ is the non-linear refractive index describing the Kerr effect and $n_{4}=-2.5\times10^{-41}$ m$^{4}$.W$^{-2}$ \cite{BergeCQ} corresponds to the quintic term.% are the cubic and quintic non-linear refractive indices.

We considered a Gaussian input pulse of $1.7$~mJ energy, $60$~fs duration (FWHM), centered at $800$~nm, with a beam radius of $3$~mm (FWHM), slightly focused by an $f=1$m lens. 
After propagating up to $z_\mathrm{f}=2$~m, the electric field was then back-propagated back to $z=0$ by changing the sign of the $z$-derivative in Equation~(\ref{eq_CubQ}). Although the pulse temporal shape (Fig.~\ref{CubQuintic_fig}(a)) and spectrum (Fig.~\ref{CubQuintic_fig}(b)) are strongly modified by the non-linear propagation along the filamentation, the back-propagated pulse perfectly overlaps with the initial one in spite of the complex propagation dynamics (Fig.~\ref{CubQuintic_fig}(d)). Fig.~\ref{CubQuintic_fig}(c) displays the residual error $(I_{\textrm{prop}}-I_{\textrm{back}})/I_{\textrm{prop}}$, where  $I_{\textrm{prop}}$ and $I_{\textrm{back}}$ are the propagated and backpropagated pulse intensities, respectively. This error does not exceed 0.06\%. This recovery error keeps minimal as soon as the numerical resolution is sufficient for the simulation to be stable.

\section{Reversibility of lossy filamentation models}

A more realistic model for filamentation can be written as
\begin{equation}
\partial_{z}\widetilde{\varepsilon}= \mathrm{i}(k_{z}-\dfrac{\omega}{v_\mathrm{g}}) \widetilde{\varepsilon} + \dfrac{\omega}{c^{2}k_{z}} \bigg[\textrm{i}\omega \Big(
n_{2} \widetilde{|\varepsilon|^{2}\varepsilon} + \Delta \widetilde{n_{\textrm{r} (t)} \varepsilon} \Big)
-\dfrac{e^{2}}{2\epsilon_{0} m_\mathrm{e}} \tau(\omega)\ \widetilde{\rho \ \varepsilon} \bigg] - \widetilde{L[\varepsilon]},
\label{eq_UPPE}
\end{equation}

Here, $m_\mathrm{e}$ is the electron mass, and $\epsilon_0$ the permittivity of vacuum. The Raman term $\Delta n_{r}(t)=\int_{-\infty}^{t} R(t') |\varepsilon(t-t')|^{2} \mathrm{d}t'$ \cite{CouaironM07} corresponds to the modification of the refractive index by molecular alignment \cite{alignment}, {calculated with a response function $R$ corresponding to the perturbation theory \cite{Morgen,Lin}, implemented as detailed in \cite{berti_ali}}. In this equation, we have introduced the quantity $\tau(\omega) = (\nu_{\mathrm{en}}+ \mathrm{i}\omega) / (\nu_{\mathrm{en}}^{2}+\omega^{2})$, where $\nu_{\mathrm{en}}$ is the collision frequency between free electrons and neutrals atoms. The last term in equation~(\ref{eq_UPPE}) accounts for plasma losses, and is calculated using
\begin{equation}
L[\varepsilon]=\frac{U_{\mathrm{i}}W(|\varepsilon|^{2})}{2|\varepsilon|^{2}}(\rho_{\mathrm{at}}-\rho)\varepsilon.
\end{equation}

When neglecting recombination of electrons over the $\lesssim$~100 fs pulse durations at stake in the present work, the free-electron density~$\rho$ is calculated with the multiphoton ionization apprimation : 
\begin{equation}
	\rho(t) = \rho_{at} \int_{-\infty}^{t} \sigma_{K}|\varepsilon|^{2K} \mathrm{d}t
\end{equation}

{where $\rho_{at}$ is the gas density. The best fit of the Keldysh-PPT (Perelomov, Popov, Terent'ev) theory of ionization \cite{Kasparian} is given by the multiphoton ionization cross section $\sigma_K = 1.5849 \times 10^{-154} \mathrm{m}^{2}\mathrm{W}^{-1}\mathrm{s}^{-1}$ (resp. $\sigma_{K}=7.9057\times 10^{-124}\mathrm{m}^{2}\mathrm{W}^{-1}\mathrm{s}^{-1}$) and $K = 9.26$ (resp. 7.5) in argon (resp. nitrogen) \cite{KaspaSC2000}}.

\begin{figure}[t!]%bhp]
	\centering
		\includegraphics[width=1.0\columnwidth]{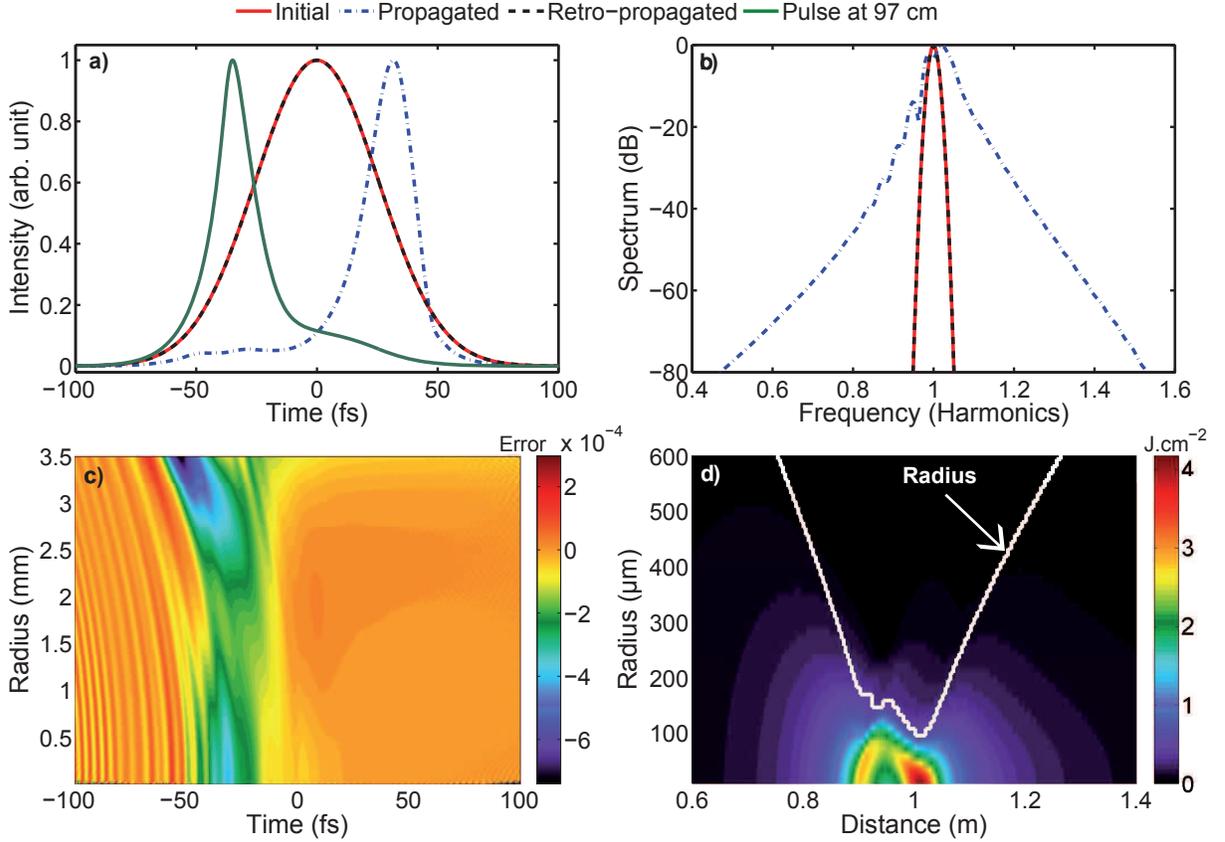}			   
		\caption{Back-propagation of a laser filament in argon when including ionization in the model. (a) On-axis intensity and (b) spectrum of the  initial (red solid line), propagated (blue dash-dotted line), back-propagated  pulses (black dashed line) and at $z=97$~cm (green solid line); (c) Relative error on intensity  between the initial and back-propagated pulses; (d) Fluence distribution along the propagation distance. The solid white line indicates the beam radius.}
	\label{Argon_fig}
\end{figure}

We {first isolated} the role of losses associated with ionization by simulating the filamentation in argon, an atomic gas in which the Raman term {is zero. To allow a comparison of the error levels,} we used the same initial conditions as for the cubic-quintic case discussed above, except for an energy of 1.5 mJ. Consistent with the introduction of plasma that induces a non-instantaneous response that couples temporal slices of the pulse with each other, the pulse propagation dynamics is more complex. In particular, a refocusing cycle is clearly visible (Fig.~\ref{Argon_fig} (d)), while pulse splitting is observed in the temporal domain in the filamentation region. In spite of these effects, both back-propagated pulse shape and spectrum are still almost indistinguishable from their initial counterparts (Fig.~\ref{Argon_fig}(a) and  Fig.~\ref{Argon_fig}(b)). 

The residual numerical error between them stays below $0.08$\% (Fig.~\ref{Argon_fig} (c)) in spite of a more complex model than the cubic-quintic model. The temporal asymmetry of the error can be directly related with the plasma generation. The largest error corresponds to the propagation distance where the intensity is maximal ($z$ = 97 cm, see Fig.~\ref{Argon_fig} (a)), hence where the plasma generation is the most efficient. 

\begin{figure}[t!]%bhp]
	\centerline{
		\includegraphics[width=1.0\columnwidth]{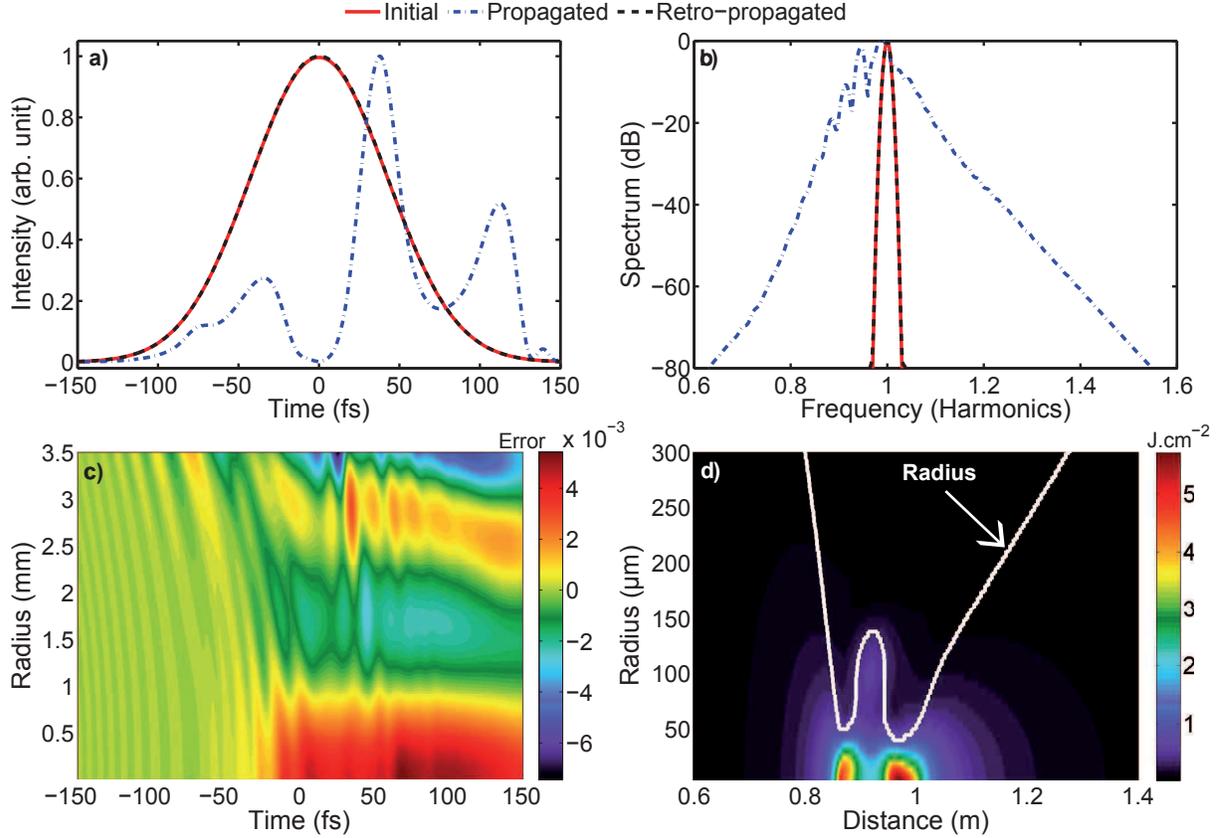}	
			   }	
		\caption{Back-propagation of a laser filament in nitrogen, taking into account ionization as well as molecular alignment. (a) On-axis intensity and (b) spectrum of the  initial (red solid line), propagated (blue dash-dotted line) and back-propagated pulses (black dashed line); (c) Relative error on intensity  between the initial and back-propagated pulses; (d) Fluence distribution along the propagation distance. The solid white line indicates the beam radius.}
	\label{N2_4bar}
\end{figure}

We then considered an additional time-retarded effect, namely molecular alignment~\cite{alignment}, by considering nitrogen at a pressure of 4~bar. This pressure was chosen to maximize the Raman contribution. 
Figure~\ref{N2_4bar} shows the result of propagation followed by back-propagation over $2$~m. The  pulse duration was stretched to $100$~fs in order to allow the Raman effect to become significant. The pulse energy was limited to $0.5$~mJ, so as to remain below the multi-filamentation regime that would be incompatible with the radial symmetry of our code.
The consideration of the Raman effect further complexifies the propagation dynamics, resulting in a very clear refocusing cycle  visible on both beam diameter and on-axis fluence (Fig.~\ref{N2_4bar} (d)). The associated pulse splitting is so strong that it results in a triple pulse that is retained even after the pulse has propagated 1~m beyond the filamenting region (Panel a). Correspondingly, its spectral counterpart is pretty much structured around the fundamental wavelength.
Again, in spite of the more complex model, one observes a very good overlap between the back-propagated pulse and the initial condition. Furthermore, the marked intra-pulse asymmetry of the numerical error plot (Fig. \ref{N2_4bar}(c)), can be associated with the asymmetry of the non-instantaneous process.

\section{Discussion}

The simulations presented above have been performed with a standard numerical propagation model, that yields results typical of other filamentation simulations in similar conditions \cite{BraunKLDSM1995,CouaironM07,Chin,Kasparian,Gavish}. For example, we obtain a peak electron densities of 5.4 $\times 10^{16}$ cm$^{-3}$ and 1.7 $\times 10^{17}$ cm$^{-3}$, inducing 8~\% and 10~\% overall energy loss in the case of the simulations in argon and nitrogen, respectively.

However, we checked that the reversibility of filamentation does not depend on the particular filamentation model considered, as long as the non-linear terms ($f$ in Eq. (\ref{eqn:NLS_CQ}) and following) depend on the electric field only vial the intensity, i.e., via $|\varepsilon|^2$. In particular, we checked that they are immune against the modeling of the propagation by the non-linear Schr\"odinger equqtion (NLSE), the consideration of the higher-order Kerr effect \cite{Bejot,B'ejHLKWF2011a}, or of alternative ionization models like PPT-ADK \cite{Kasparian}. Furthermore, the present work focus on the propagation equations expressed in terms of the field envelope, but the propagation equations of the real field can be shown to be similarly reversible.

Filament reversibility is therefore a general property, that may appear as a paradox with regard to the intensity clamping \cite{KaspaSC2000,DaiglJHWKRBC2010}, that apparently induce losses of memory of the initial conditions. This apparent paradox can be solved by considering that the electric field bears a much richer information than can be measured when considering one single observable, e.g.,  the intensity or spectral phase in the filament center. Rather, even if such unique observable seems to loose memory, the information redistribution in space and over all observables ensures that the information required to back-propagate the pulse is in fact conserved. 
Indeed, it is well-known that the filaments are not an isolated system, but are in strong interaction with the surrounding photon bath \cite{CourvBKSMYW2003,LiuTAGBC2005,Scheller2014}. 
The self-healing nature of filamentation illustrates that the photon bath alone contains sufficient information to reconstruct a blocked filament \cite{CourvBKSMYW2003}, via the spatial distribution of the spectrum, spectral phase, and transverse wavenumber spectrum.

\section{Conclusion}
As a conclusion, our work demonstrates the unexpected reversibility of filamentation, even when plasma losses are taken into account. It illustrates the redistribution of information within the whole filament-photon bath system. This result constitute a clear evidence that the actual time-reversibility of real physical systems can expand beyond the domain of formally well-posed inverse problems. It may open the way to pulse design for controlling or optimizing filaments in view of specific applications.

\section*{Acknowledgments}
We acknowledge financial support by the European Research Council Advanced Grant ``Filatmo''. Discussions with Pierre B\'ejot were very rewarding and the assistance of Michel Moret was highly appreciated.

\bibliographystyle{unsrt}

\end{document}